\documentclass[final,5p,times,numbers]{elsarticle}
\usepackage{amsmath,amssymb,amsfonts}
\usepackage{graphicx}
\graphicspath{{Pictures/}}
\usepackage{hyperref}
\hypersetup{pdfnewwindow=true, colorlinks=true, linkcolor=blue, anchorcolor=blue, citecolor=red, filecolor=blue, menucolor=blue, urlcolor=magenta}
\usepackage{cleveref}
\usepackage{color}
\usepackage{float}
\usepackage{soul}
\usepackage{ulem}
\usepackage{braket}
\usepackage{enumitem}
\usepackage{marginnote}
\usepackage{orcidlink}

\usepackage{microtype}
\usepackage{xurl}

\newcommand{\onlinecite}[1]{\citealp{#1}}
\newenvironment{widetext}{}{}

\newcommand{\blue}[1] {{\color{black}#1}}

\newcommand{\dg}{\ensuremath{g}}

\journal{Physics Letters A}
\biboptions{sort&compress}

\begin{document}

\begin{frontmatter}

\title{Entanglement and reciprocal dynamics of the Barnett and Einstein-de Haas responses in a quantum spin-rotor model}

\author[greifswald]{Saikat Banerjee\,\orcidlink{0000-0002-3397-0308}}
\ead{saikat.banerjee@uni-greifswald.de}
\author[greifswald,erlangen]{Holger Fehske\,\orcidlink{0000-0003-2146-8203}}
\ead{fehske@physik.uni-greifswald.de}

\address[greifswald]{Institute of Physics, University of Greifswald, Felix-Hausdorff-Stra{\ss}e 6, 17489 Greifswald, Germany}
\address[erlangen]{Erlangen National High Performance Computing Center, Friedrich-Alexander-Universit\"{a}t Erlangen-N\"{u}rnberg, 91058 Erlangen, Germany}

%------------------
\begin{abstract}
We study the reciprocal Barnett and Einstein--de Haas effects in an exactly solvable quantum spin--rotor model that conserves the total angular momentum. For a single spin-$1/2$, the model describes rotation-dependent spin polarization, coherent transfer of angular momentum between the spin and the rotor, and the accompanying spin--rotor entanglement. At a recoil-compensation point, all angular-momentum sectors become synchronized, allowing complete spin reversal and a one-quantum shift of an arbitrary rotor wave packet. 
%We then extend the model to two spins coupled to the same rotor. The rotor can now mediate entanglement between two distinct spins rather than only becoming entangled with the spin itself. 
\blue{We then extend the model to two spins coupled to the same rotor, where the shared rotational degree of freedom participates dynamically in the generation of spin--spin entanglement.} The singlet forms a dark state, while the triplet sector remains coupled to the mechanical motion. At a special parameter point, the rotor returns to its original dynamical branch and disentangles from the spins, leaving behind a maximally entangled two-spin state locally equivalent to a Bell state.
\end{abstract}
%------------------

\end{frontmatter}

%--------------------------
\section{Introduction \label{sec.I}} 
%--------------------------

Mechanical rotation and magnetization are linked by the exchange of angular momentum. In its conventional form, the Barnett effect~\cite{PhysRev.6.239,RevModPhys.7.129} refers to the magnetization induced by rotating a body, whereas the reciprocal Einstein--de Haas effect~\cite{PhysRevSeriesI.26.248,Viktor_1979} converts a change of spin angular momentum into mechanical motion. These reciprocal responses have been studied in magnetic cantilevers, many-body spin systems, single-molecule magnets, and spin-current and magnetic-resonance experiments~\cite{PhysRevB.79.104410,PhysRevB.99.064428,Ganzhorn2016,PhysRevLett.106.076601,Kurimune2020}. On the Barnett side, mechanical rotation can often be represented by an effective magnetic field acting on the spin degrees of freedom, leading to rotation-induced spin polarization and related inertial phenomena~\cite{PhysRevB.84.104410,PhysRevB.92.174424,PhysRevB.103.174308,Chudo2021,Maekawa2023,PhysRevB.111.L060403,Zhang2016}. A natural question is how the Barnett response and its reciprocal angular-momentum transfer appear when the mechanical rotation itself is treated as a quantum degree of freedom.

A quantum rotor is qualitatively different from the harmonic mechanical modes commonly used in spin--boson and optomechanical models. Its angular coordinate $\phi$ is compact, its angular momentum $L_z$ has a discrete spectrum, and the operators $e^{\pm i\phi}$ change the mechanical angular momentum by one quantum. The importance of total-angular-momentum conservation in magnetic-moment tunneling was emphasized in Ref.~\onlinecite{PhysRevLett.72.3433}, while rotational states of freely rotating nanomagnets and more general quantum magnetic rotors were studied in Refs.~\onlinecite{ChudnovskyGaranin2010,PhysRevB.93.054427}. Spin-controlled nanoparticle rotation~\cite{PhysRevB.104.134310}, quantum Einstein--de Haas dynamics in nanomechanical and nanorotor systems~\cite{Ganzhorn2016,Izumida2022}, and recently proposed levitated quantum spin--rotor platforms~\cite{zjpt-whpf} further illustrate the growing interest in treating spin and mechanical angular momentum on the same quantum footing. %In contrast to Dicke-type models~\cite{PhysRevA.88.043835,PhysRevA.85.043821,Kirton2019}, the problem considered here does not rely on a soft bosonic mode or a symmetry-breaking phase transition. Instead, the compact rotor allows mechanical recoil and total-angular-momentum conservation to be followed explicitly at the level of individual angular-momentum quanta.
\blue{Although the symmetric collective-coupling limit considered below is structurally reminiscent of Dicke-type models~\cite{PhysRevA.88.043835,PhysRevA.85.043821,Kirton2019}, the present problem differs in an essential way: the shared dynamical degree of freedom is a compact rotor rather than a harmonic bosonic mode. The operators $e^{\pm i\phi}$ therefore implement discrete mechanical recoil, allowing total-angular-momentum conservation to be followed explicitly at the level of individual angular-momentum quanta.}

Mechanical degrees of freedom are also of interest as quantum mediators. Vibrational modes have been proposed to generate effective interactions between otherwise separated spins and to create collective spin entanglement~\cite{Bennett2013,Xia2016}, and mechanically mediated interactions have recently been explored as a route toward programmable spin-qubit networks~\cite{Fung2024}. These approaches are mainly based on harmonic vibrational modes and platform-specific spin--phonon couplings. This raises a complementary question in the present gyromagnetic setting: can a quantum rotor mediate entanglement between two spins through the same angular-momentum exchange that underlies the Einstein--de Haas effect, and can the rotor subsequently separate from the spins while leaving the spin pair entangled?

Here we answer these questions within an exactly solvable spin--rotor model. For a single spin-$1/2$, a rotor-phase-dependent transverse coupling changes the spin and mechanical angular momenta by equal and opposite quanta and therefore conserves $J_z=L_z+S_z$. 
%The resulting fixed-$J_z$ sectors give a quantum-rotor description of the Barnett response and, out of equilibrium, describe coherent Einstein--de Haas transfer and spin--rotor entanglement. 
\blue{The resulting fixed-$J_z$ sectors provide a quantum-rotor counterpart of the Barnett response, which reduces to the usual rotation-induced effective-field description in the semiclassical limit; out of equilibrium, the same model realizes a coherent quantum analogue of Einstein--de Haas angular-momentum transfer and spin--rotor entanglement.} At a recoil-compensation point, the dynamics becomes synchronized across all angular-momentum sectors, leading to complete spin reversal and a one-quantum shift of an arbitrary initial rotor wave packet. We then extend the model to two spins coupled to the same rotor. The two-spin Hilbert space separates into a dark singlet and a rotor-active triplet sector. 
%This structure allows an initially separable spin pair to acquire entanglement through a coherent rotor-mediated cycle.
\blue{This structure allows an initially separable spin pair to become entangled through a coherent cycle in which the rotor participates dynamically and subsequently disentangles.} At an analytically determined parameter point, the rotor returns to a factorized state while the two spins are left maximally entangled, in a state locally equivalent to a Bell state. 
%The model therefore provides an exactly solvable connection between reciprocal gyromagnetic dynamics, mechanical recoil, and rotor-mediated quantum entanglement. 
\blue{The model is intended as a minimal coherent benchmark for an axially constrained quantum rotor rather than as a microscopic description of a specific molecular or nanomechanical device.}

%--------------------------
\section{Model \label{sec.II}}
%--------------------------

We consider an effective spin-$1/2$ particle coupled to a planar quantum rotor,
%----------------------------
\begin{equation}\label{eq.1}
{\cal H} = \frac{L_z^2}{2I} + \frac{\Delta}{2} \left( S_{+}e^{-i\phi} + S_{-}e^{i\phi} \right) + \dg L_z S_z. 
\end{equation}
%----------------------------
where $L_z$ is the rotor angular momentum, $I$ is the moment of inertia, $\Delta$ is a transverse spin splitting, and $\dg$ denotes the spin--rotation coupling. Here, $\phi$ is the angular coordinate of the rotor, $S_{\pm}=S_x\pm iS_y$, and we set $\hbar=1$. In the angular-momentum basis, $L_z|m\rangle=m|m\rangle$ and $e^{\pm i\phi}|m\rangle=|m\pm1\rangle$. The first term in Eq.~\eqref{eq.1} is the kinetic energy of the rotor, the second describes a transverse spin-flip process whose laboratory-frame form depends on the rotor orientation, and the third represents a longitudinal coupling between the mechanical and spin angular momenta.

The relevant angular-momentum commutation relations are
%-------------------------------
\begin{equation}\label{eq.2}
[L_z,e^{\pm i\phi}] = \pm e^{\pm i\phi}, \qquad [S_z,S_{\pm}] = \pm S_{\pm}.
\end{equation}
%-------------------------------
The Hamiltonian may be viewed as an effective local description of a spin-orbit-coupled moment embedded in a rotating ligand or molecular environment, for example in octahedral or tetrahedral coordination units or in gyroscope-like metal complexes~\cite{Swart2016,Ehnbom2021,Fry2013,Shotwell2025,Setaka2013}. In such a setting, the local axis associated with the transverse splitting is fixed in the body frame and therefore rotates through the angle $\phi$ in the laboratory frame. Indeed, the transverse term may equivalently be written as
\[
\frac{1}{2} \left( S_{+}e^{-i\phi} + S_{-}e^{i\phi} \right) = S_x\cos\phi+S_y\sin\phi.
\]
Accordingly, $\Delta$ should be interpreted as an effective transverse splitting in the body-fixed frame, while $\dg$ represents an additional longitudinal spin-rotation coupling. \blue{More generally, the spin--rotation interaction may be written in tensor form as 
$$
{\cal H}_{\rm sr}=\mathbf{L}\cdot\boldsymbol{\Gamma}\cdot\mathbf{S},
$$ 
where the tensor $\boldsymbol{\Gamma}$ is determined by the microscopic spin--orbit and local crystal-field environment. When the rotational motion is restricted to a single axial degree of freedom about $z$, retaining the
corresponding longitudinal component reduces this coupling to $\dg L_zS_z$. Thus, Eq.~\eqref{eq.1} should be regarded as the effective low-energy Hamiltonian obtained after restricting both the local spin manifold and the mechanical motion to their relevant low-energy degrees of freedom.} Both are effective low-energy parameters determined by crystal-field, spin-orbit, and local-environment effects rather than bare microscopic couplings. Equation~\eqref{eq.2} makes the angular-momentum exchange explicit. The operator $S_{+}e^{-i\phi}$ raises the spin angular momentum and lowers the rotor angular momentum by one quantum, whereas $S_{-}e^{i\phi}$ produces the reverse process. Defining the total angular momentum as $J_z=L_z+S_z$, one finds
%-------------------------------
\begin{equation}\label{eq.3}
\begin{aligned}
[J_z,S_{\pm}e^{\mp i\phi}] &= [S_z,S_{\pm}]e^{\mp i\phi} + S_{\pm}[L_z,e^{\mp i\phi}] \\
&=0,
\end{aligned}
\end{equation}
%-------------------------------
and consequently
%-------------------------------
\begin{equation}\label{eq.4}
[{\cal H},J_z]=0.
\end{equation}
%-------------------------------
Thus, $J_z$ is conserved. The Hamiltonian can consequently be decomposed into independent sectors of fixed total angular momentum and solved exactly, while retaining the essential ingredients required to describe rotation-dependent Barnett response and reciprocal Einstein--de Haas angular-momentum transfer. This effective interpretation also provides a schematic connection to more realistic local-moment and spin--lattice studies in correlated materials~\cite{PhysRevB.105.L180414,Kumar2022,Kim2024,Schaffer_2016,Paschen2021,Chudo2021}. \blue{At this point, we would like to emphasize that Eq.~\eqref{eq.1} is intended as a minimal effective spin--rotor model rather than as a microscopic description of a specific molecular or nanomechanical device. The planar rotor represents a single axially constrained rotational or torsional degree of freedom. A freely rotating three-dimensional rigid body contains additional rotational degrees of freedom and, in general, does not retain the simple fixed-$J_z$ block structure exploited below. Accordingly, the effective parameters $\Delta$, $\dg$, and $I$ must be determined for any specific microscopic realization.}

%-----------------------------------------------------------------
\subsection{Exact solution and Barnett response \label{sec.II.A}}
%-----------------------------------------------------------------

Since the Hamiltonian in Eq.~\eqref{eq.1} conserves the total angular momentum $J_z=L_z+S_z$, rather than the mechanical angular momentum $L_z$ and spin angular momentum $S_z$ separately, the Hilbert space can be decomposed into independent sectors labelled by the eigenvalue $j$ of $J_z$. For a spin-$1/2$, $j$ is half-integer, and each fixed-$j$ sector is spanned by the two states
%----------------------------
\begin{equation}\label{eq.5}
\ket{j,\uparrow} \equiv  \ket{j-\frac{1}{2}}\otimes\ket{\uparrow}, \qquad \ket{j,\downarrow} \equiv \ket{j+\frac{1}{2}}\otimes\ket{\downarrow}.
\end{equation}
%----------------------------
Both states carry the same total angular momentum $j$, since the difference in their spin angular momenta is compensated by a one-quantum difference in the mechanical angular momentum. The phase-dependent transverse coupling connects precisely these two states,
%----------------------------
\begin{equation}\label{eq.6}
S_{+}e^{-i\phi}\ket{j,\downarrow} = \ket{j,\uparrow}, \qquad S_{-}e^{i\phi}\ket{j,\uparrow} = \ket{j,\downarrow}.
\end{equation}
%----------------------------
In the basis $\{\ket{j,\uparrow},\ket{j,\downarrow}\}$, the Hamiltonian therefore reduces to the $2\times2$ matrix written in a compact form as
%----------------------------
\begin{equation}\label{eq.8}
{\cal H}_{j} = \varepsilon_j^{(0)}\,\mathbb{I}_2 + \frac{1}{2}\left( \Delta\sigma_x + \delta_j\sigma_z \right), 
\end{equation}
%----------------------------
where
%----------------------------
\begin{equation}\label{eq.9}
\varepsilon_j^{(0)} = \frac{j^2}{2I} + \frac{1}{8I} - \frac{\dg}{4}, \qquad \delta_j = j\left(\dg-\frac{1}{I}\right).
\end{equation}
%----------------------------

%----------------------------------------------------------------------
%\subsubsection{Transformation to body-fixed frame \label{sec.II.A.1}}
%----------------------------------------------------------------------

The origin of the two contributions to $\delta_j$ can also be understood by transforming to the body-fixed frame. In the angular representation, we introduce $U=e^{-i\phi S_z}$ as a unitary operator to transform the Hamiltonian as $\widetilde{\cal H} =
U^{\dag}{\cal H}U$. Using $U^{\dag}L_z U = L_z-S_z$ and 
\[
U^{\dag} \left(S_{+}e^{-i\phi}+S_{-}e^{i\phi}\right)U = S_{+}+S_{-},
\]
the transformed Hamiltonian becomes
%-------------------------------
\begin{equation}\label{eq.body_frame}
\widetilde{\cal H} = \frac{L_z^2}{2I} + \Delta S_x + \left( \dg-\frac{1}{I} \right)L_zS_z + \left( \frac{1}{2I}-\dg \right)S_z^2. 
\end{equation}
%-------------------------------
For spin-$1/2$, $S_z^2=1/4$, so the last term in Eq.~\eqref{eq.body_frame} is an additive constant. Moreover, $U^\dag J_z U = L_z$~\footnote{For spin-$1/2$, ${\cal U}(\phi+2\pi)=-{\cal U}(\phi)$. Consequently, the transformed wave function obeys antiperiodic boundary conditions, and the eigenvalues of the transformed $L_z$ are the half-integer values $j$ of the conserved total angular momentum.}, showing that the transformed angular-momentum carries the conserved sector label $j$. The longitudinal term in Eq.~\eqref{eq.body_frame} therefore reproduces the detuning $\delta_j=j(\dg-1/I)$ obtained in Eq.~\eqref{eq.9}. At the compensation condition $\dg I=1$, the transformed Hamiltonian reduces to
%-------------------------------
\begin{equation}\label{eq.100}
\widetilde{\cal H} \big|_{\dg I=1} = \frac{L_z^2}{2I} +\Delta S_x - \frac{1}{8I}. 
\end{equation}
%-------------------------------
Thus, the spin dynamics becomes independent of the conserved angular-momentum sector in the body-fixed frame, explaining why all fixed-$J_z$ sectors become resonant with the same frequency. We return later to the coherent spin and rotor dynamics at this special compensation point.

%-----------------------------------------------------------------
%\subsubsection{Spectrum and Barnett response \label{sec.II.A.2}}
%-----------------------------------------------------------------

\begin{figure}[b]
\centering
\includegraphics[width=1.0\linewidth]{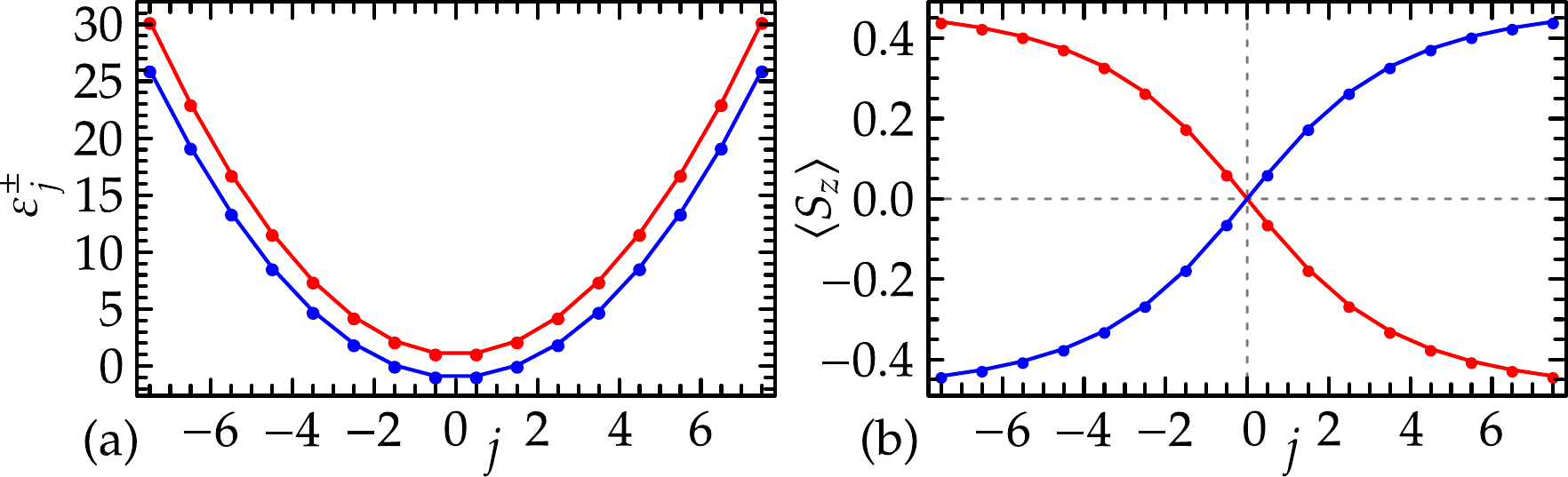}
\caption{(a) Exact energy branches $\varepsilon_j^{\pm}$ as functions of the conserved total angular momentum $j$. The markers denote the allowed half-integer values of $j$, while the lines are guides to the eye. (b) Corresponding spin polarizations $\langle S_z\rangle_{j,\pm}$. The markers show the exact quantum-rotor results, while the solid curves represent the continuous semiclassical result obtained from ${\cal H}_{\rm s}^{\rm rot} = \Delta S_x+(\dg I-1)\omega S_z$ after identifying $\omega=j/I$. The parameters are $\Delta I=2$ and $\dg I=0.5$.}
\label{fig:Fig1}
\end{figure}

Returning to the fixed-$j$ representation, $\delta_j$ is the effective longitudinal detuning. As made explicit by the body-fixed-frame transformation, it contains the spin--rotation contribution $\dg j$ and the mechanical-recoil contribution $-j/I$. The exact eigenenergies are
%----------------------------
\begin{equation}\label{eq.10}
\varepsilon_{j}^{\pm} = \varepsilon_j^{(0)} \pm \frac{1}{2}\Omega_j, \qquad \Omega_j = \sqrt{\Delta^2+\delta_j^2}.
\end{equation}
%----------------------------
Thus, each conserved total angular-momentum sector is equivalent to a two-level spin problem in the effective field
%----------------------------
\begin{equation}\label{eq.11}
{\bf B}_{j}^{\rm eff} = \Delta\,\hat{\bf x} + \delta_j\,\hat{\bf z}. 
\end{equation}
%----------------------------
Introducing the angle $\theta_j$ through $\cos\theta_j=\delta_j/\Omega_j$ and $\sin\theta_j=\Delta/\Omega_j$, the spin polarization in the two eigenstates is 
%----------------------------
\begin{equation}\label{eq.12}
\langle S_z\rangle_{j,\pm} = \pm\frac{1}{2} \frac{\delta_j}{\Omega_j}. 
\end{equation}
%----------------------------
Equations~\eqref{eq.10} and \eqref{eq.12} show that both the energy splitting and the spin polarization depend explicitly on the conserved total-angular-momentum sector. In the regime where the rotor wave packet is centered around a large angular momentum with small relative fluctuations, this sector dependence provides the quantum-rotor analogue of the Barnett response. In particular, when the mechanical-recoil correction is small, $1/I\ll|\dg|$, one recovers the effective longitudinal Barnett field $\delta_j\simeq\dg j$. The corresponding sector-dependent energy branches and spin polarizations are shown in Fig.~\ref{fig:Fig1}.

%----------------------------------
%----------------------------------

For a rotor wave packet narrowly centered around a large angular momentum $j_0$, with relative fluctuations $\sigma_j/|j_0|\ll1$, the sector-dependent detuning $\delta_j$ may be approximated by $\delta_{j_0}$. The spin then experiences an effectively static longitudinal field, recovering the familiar semiclassical description of the Barnett response.

This connection can be made more explicit by considering a rotor with a well-defined angular velocity, $\phi(t)=\omega t$, and mechanical angular momentum $L_z\simeq I\omega$. In this limit, the spin-dependent part of Eq.~\eqref{eq.1} becomes
%----------------------------
\begin{equation}\label{eq.13}
{\cal H}_{\rm s}(t) = \Delta \left[ S_x\cos(\omega t) + S_y\sin(\omega t) \right] + \dg I\omega S_z. 
\end{equation}
%----------------------------
Upon transforming to the frame corotating with the rotor, the spin Hamiltonian takes the time-independent form
%----------------------------
\begin{equation}\label{eq.14}
{\cal H}_{\rm s}^{\rm rot} = \Delta S_x + \left(\dg I-1\right)\omega S_z. 
\end{equation}
%----------------------------
The longitudinal term therefore defines an effective Barnett field proportional to the angular velocity. For a quantum rotor wave packet narrowly centered around a large $j_0$, one has $j_0/I\simeq\omega$, and hence $\delta_{j_0}\simeq(\dg I-1)\omega$, in agreement with Eq.~\eqref{eq.9}. The two contributions to the effective longitudinal field have a transparent origin: $\dg I\omega$ arises from the explicit longitudinal spin--rotation coupling, whereas $-\omega$ is the kinematic contribution generated by the transformation to the corotating frame. In the fully quantum description, the latter corresponds to the recoil correction arising from the difference between the kinetic energies of the two coupled rotor states. \blue{In this sense, the sector-dependent polarization of the quantum rotor provides the quantum counterpart of the conventional Barnett response: in the semiclassical limit it reduces to the spin polarization induced by an effective field proportional to the angular velocity. We emphasize, however, that a fixed-$j$ eigenstate is not itself a dynamical preparation protocol for the macroscopic Barnett effect.}

%------------------------------------------------------------------
\subsection{Coherent Einstein--de Haas dynamics \label{sec.II.B}}
%------------------------------------------------------------------

%The same fixed-$j$ decomposition provides a direct description of the reciprocal Einstein--de Haas response. 
\blue{The same fixed-$j$ decomposition provides a coherent quantum analogue of the reciprocal Einstein--de Haas response.}
To display the transfer of spin angular momentum to the mechanical rotor, we consider the initial state
%----------------------------
\begin{equation}\label{eq:initial_state}
\ket{\psi_j(0)} = \ket{j,\uparrow} = \ket{j-\frac{1}{2}}\otimes\ket{\uparrow}.
\end{equation}
%----------------------------
The phase-dependent transverse term couples this state to $\ket{j,\downarrow} = \ket{j+\frac{1}{2}}\otimes\ket{\downarrow}$. A spin flip therefore increases the rotor angular momentum by one quantum while decreasing the spin angular momentum by the same amount. Using the block Hamiltonian in Eq.~\eqref{eq.8}, the exact time-evolved state is
%----------------------------
\begin{equation}\label{eq:time_evolved_state}
\begin{aligned}
\ket{\psi_j(t)} = e^{-i\varepsilon_j^{(0)}t} \Bigg[ & \left( \cos\frac{\Omega_jt}{2} - i\frac{\delta_j}{\Omega_j} \sin\frac{\Omega_jt}{2} \right) \ket{j,\uparrow} \\
&
- i\frac{\Delta}{\Omega_j} \sin\frac{\Omega_jt}{2} \ket{j,\downarrow} \Bigg]. 
\end{aligned}
\end{equation}
%----------------------------
The probability for the spin to flip while transferring one quantum of angular momentum to the rotor is therefore
%----------------------------
\begin{equation}\label{eq:transfer_probability}
P_j(t) = \frac{\Delta^2}{\Omega_j^2} \sin^2\left(\frac{\Omega_jt}{2}\right).
\end{equation}
%----------------------------
The corresponding expectation values of the spin, mechanical, and total angular momenta are
%----------------------------
\begin{equation}\label{eq:angular_momentum_expectation}
\begin{gathered}
\langle S_z(t)\rangle = \frac{1}{2}-P_j(t), \\
\langle L_z(t)\rangle = j-\frac{1}{2}+P_j(t), \\
\langle J_z(t)\rangle = j.
\end{gathered}
\end{equation}
%----------------------------
Consequently, the changes in the spin and rotor angular momenta satisfy
%----------------------------
\begin{equation}\label{eq:angular_momentum_transfer}
\langle L_z(t)\rangle-\langle L_z(0)\rangle = - \left[ \langle S_z(t)\rangle-\langle S_z(0)\rangle \right] = P_j(t).
\end{equation}
%----------------------------
Eq.~\eqref{eq:angular_momentum_transfer} is the dynamical manifestation of total angular-momentum conservation. In particular, the populations of the neighboring rotor states $\ket{j-\frac{1}{2}}$ and $\ket{j+\frac{1}{2}}$ evolve explicitly in time. %The resulting coherent exchange provides a minimal quantum description of the Einstein--de Haas response.
\blue{The resulting coherent and reversible exchange provides a minimal quantum analogue of the Einstein--de Haas angular-momentum-transfer process. The conventional macroscopic effect generally also involves relaxation and dissipation, which are absent from the present closed-system model.}

Mathematically, Eq.~\eqref{eq:transfer_probability} has the form of a detuned two-level Rabi oscillation. Its gyromagnetic content follows from the nature of the two coupled states, which differ by one quantum of spin angular momentum and an oppositely directed quantum of mechanical angular momentum. \blue{We therefore do not regard the two-level Rabi structure itself as a new result. Rather, its role here is to provide an exactly solvable spin--rotor realization in which the two coupled states differ by one quantum of mechanical angular momentum, so that the Rabi dynamics directly describes coherent spin--mechanical angular-momentum transfer.} The maximum transfer probability within the fixed-$j$ sector is $P_j^{\rm max} = \tfrac{\Delta^2}{\Delta^2+\delta_j^2}$. The sector-dependent detuning also identifies a special recoil-compensation point. When $\dg I=1$, one has $\delta_j=0$ and $\Omega_j=\Delta$ for every allowed value of $j$. As also seen from the body-fixed-frame Hamiltonian, the spin dynamics then becomes independent of the conserved total-angular-momentum sector. Consequently, all sectors undergo resonant spin--rotor oscillations with the same frequency. 

\blue{The exact condition $\dg I=1$ is a special compensation point, but the transfer dynamics varies smoothly in its vicinity. To quantify this, we write
%-------------------------------
\begin{equation}\label{eq.1c.1}
\epsilon=\dg I-1, \qquad d=\Delta I .    
\end{equation}
%-------------------------------
Then $\delta_j=\frac{j\epsilon}{I}$, and the maximum transfer probability becomes
%-------------------------------
\begin{equation}\label{eq.1c.2}
P_j^{\rm max}  = \frac{d^2}{d^2+j^2\epsilon^2} = 1-\frac{j^2}{d^2}\epsilon^2 +\mathcal{O}(\epsilon^4).    
\end{equation}
%-------------------------------
Likewise, the oscillation frequency is
%-------------------------------
\begin{equation}\label{eq.1c.3}
\Omega_j = \Delta \left[ 1+\frac{j^2}{2d^2}\epsilon^2 + \mathcal{O}(\epsilon^4)\right].    
\end{equation}
%-------------------------------
Thus, both the reduction of the maximum transfer probability and the sector-dependent correction to the oscillation frequency are quadratic in a small deviation from the compensation point. For a rotor wave packet, the compensated dynamics therefore remains a good approximation provided $|j\epsilon|\ll d$ for the angular-momentum sectors with appreciable occupation. We emphasize, however, that exact compensation is not expected to be generic: $\dg$ and $I$ are effective microscopic and mechanical parameters whose values depend on the particular realization. The condition $\dg I=1$ should therefore be understood as identifying an analytically special synchronization point, together with a finite regime around it in which the transfer dynamics remains close to the compensated limit.}

To demonstrate the resulting synchronized transfer, we consider an arbitrary initial rotor state,
%----------------------------
\begin{equation}\label{eq:general_initial}
\ket{\Psi(0)} = \ket{\chi}\otimes\ket{\uparrow}, \quad \ket{\chi} = \sum_{m\in\mathbb{Z}}c_m\ket{m},
\end{equation}
%----------------------------
with $\sum_m|c_m|^2=1$. At the compensation point, its exact time evolution is
%----------------------------
\begin{equation}\label{eq:general_evolution}
\begin{aligned}
\ket{\Psi(t)}  = \sum_m c_m &  e^{-i\frac{m(m+1)}{2I}t} \Bigg[ \cos\left(\frac{\Delta t}{2}\right) \ket{m}\otimes\ket{\uparrow} \\
& - i\sin\left(\frac{\Delta t}{2}\right) \ket{m+1}\otimes\ket{\downarrow} \Bigg].
\end{aligned}
\end{equation}
%----------------------------
Defining the freely phase-evolved rotor state
%----------------------------
\begin{equation}\label{eq:chi_t}
\ket{\chi_t} = \sum_m c_me^{-i\frac{m(m+1)}{2I}t} \ket{m},
\end{equation}
%----------------------------
Eq.~\eqref{eq:general_evolution} can equivalently be written as
%----------------------------
\begin{equation}\label{eq:general_compact}
\begin{aligned}
\ket{\Psi(t)} & = \cos\left(\frac{\Delta t}{2}\right) \ket{\chi_t}\otimes\ket{\uparrow} \\
& - i\sin\left(\frac{\Delta t}{2}\right) e^{i\phi}\ket{\chi_t}\otimes\ket{\downarrow}.
\end{aligned}
\end{equation}
%----------------------------
At the transfer times
$t_{\pi}=(2n+1)\pi/\Delta$, with integer $n$, the evolution becomes
%----------------------------
\begin{equation}\label{eq:complete_transfer}
\ket{\Psi(t_{\pi})} = -i(-1)^n e^{i\phi} \ket{\chi_{t_{\pi}}} \otimes \ket{\downarrow}.
\end{equation}
%----------------------------
Thus, complete spin reversal occurs for an arbitrary initial rotor wave packet, while every angular-momentum component of the mechanical state is shifted upward by exactly one quantum. Correspondingly, $\langle S_z(t_{\pi})\rangle-\langle S_z(0)\rangle=-1$ and $\langle L_z(t_{\pi})\rangle-\langle L_z(0)\rangle=+1$, so that the total angular momentum remains conserved. %Away from the compensation condition $\dg I=1$, different total-angular-momentum sectors generally acquire different detunings and oscillation frequencies, and the universal synchronized transfer is lost.
\blue{Away from the exact compensation condition $\dg I=1$, different total-angular-momentum sectors acquire different detunings and oscillation frequencies, so the exact universal synchronization is lost; for sufficiently small deviations, however, the resulting reduction of the transfer fidelity is controlled by the quadratic corrections discussed above.}

%-----------------------------------------------------------------------------------------
\subsection{Spin--rotor entanglement and angular-momentum correlations \label{sec.II.C}}
%-----------------------------------------------------------------------------------------

The coherent Einstein--de Haas process also generates spin--rotor entanglement with respect to the laboratory-frame spin--rotor system. From Eq.~\eqref{eq:time_evolved_state}, the time-evolved state within a fixed-$j$ sector may be written as
%----------------------------
\begin{equation}\label{eq:entangled_state}
\ket{\psi_j(t)} = a_j(t) \ket{j-\frac{1}{2}}\otimes\ket{\uparrow} + b_j(t) \ket{j+\frac{1}{2}}\otimes\ket{\downarrow}, 
\end{equation}
%----------------------------
where
%----------------------------
\begin{equation}\label{eq:entangled_amplitudes}
\begin{aligned}
a_j(t) & = e^{-i\varepsilon_j^{(0)}t} \left[ \cos\left(\frac{\Omega_jt}{2}\right) - i\frac{\delta_j}{\Omega_j} \sin\left(\frac{\Omega_jt}{2}\right) \right], \\
b_j(t) & = -i e^{-i\varepsilon_j^{(0)}t} \frac{\Delta}{\Omega_j} \sin\left(\frac{\Omega_jt}{2}\right). 
\end{aligned}
\end{equation}
%----------------------------
The corresponding probabilities satisfy $|a_j(t)|^2=1-P_j(t)$ and $|b_j(t)|^2=P_j(t)$, where $P_j(t)$ is the angular-momentum-transfer probability defined in Eq.~\eqref{eq:transfer_probability}.

Within a fixed-$j$ sector, the two occupied rotor states may be regarded as the states of a logical rotor qubit,
%----------------------------
\begin{equation}\label{eq:logical_rotor_qubit}
\ket{0}_{\rm R} \equiv \ket{j-\frac{1}{2}}, \qquad \ket{1}_{\rm R} \equiv \ket{j+\frac{1}{2}}.
\end{equation}
%----------------------------
Eq.~\eqref{eq:entangled_state} is therefore, up to local phase factors, a pure two-qubit state in Schmidt form. The two components contain orthogonal spin  and rotor states whose angular momenta differ by one quantum, except when either $a_j(t)$ or $b_j(t)$ vanishes.

Tracing out the rotor gives the reduced spin density matrix
%----------------------------
\begin{equation}\label{eq:rho_spin_fixed}
\begin{aligned}
\rho_{\rm spin}^{(j)}(t) & = \operatorname{Tr}_{\rm rotor} \left[ \ket{\psi_j(t)}\bra{\psi_j(t)} \right] \\
						 & = \left[1-P_j(t)\right] \ket{\uparrow}\bra{\uparrow} + P_j(t) \ket{\downarrow}\bra{\downarrow}. 
\end{aligned}
\end{equation}
%----------------------------
Similarly, tracing out the spin gives
%----------------------------
\begin{equation}\label{eq:rho_rotor_fixed}
\begin{aligned}
\rho_{\rm rotor}^{(j)}(t) & = \operatorname{Tr}_{\rm spin} \left[ \ket{\psi_j(t)}\bra{\psi_j(t)} \right] \\
						  & = \left[1-P_j(t)\right] \ket{0}_{\rm R}\bra{0}_{\rm R} + P_j(t) \ket{1}_{\rm R}\bra{1}_{\rm R}. 
\end{aligned}
\end{equation}
%----------------------------
The two reduced density-matrices have the same nonzero eigenvalues, $1-P_j(t)$ and $P_j(t)$. Thus, the same transition probability controls the spin-flip probability, the mechanical-angular-momentum transfer, and the spin--rotor entanglement.

For a pure two-qubit state, the concurrence is defined in terms of its probability amplitudes~\cite{Wootters1998}. Applied to Eq.~\eqref{eq:entangled_state}, it takes the particularly simple form
%----------------------------
\begin{equation}\label{eq:concurrence_fixed}
{\cal C}_j(t) = 2\left|a_j(t)b_j(t)\right| = 2\sqrt{P_j(t)\left[1-P_j(t)\right]}.
\end{equation}
%----------------------------
\blue{This relation follows directly from the pure two-level Schmidt structure of the fixed-$j$ state and is therefore not specific to the spin--rotor model. In particular, the maximum at $P_j=1/2$ is the generic equal-weight condition for a two-level bipartite state.} The pure-state I-concurrence, which applies to bipartite systems of arbitrary dimensions, is defined by~\cite{Rungta2001,Glaser2003}
%----------------------------
\begin{equation}\label{eq:iconcurrence_fixed}
{\cal C}_{I,j}(t) = \sqrt{ 2\left[ 1-\operatorname{Tr}\left(\left[\rho_{\rm spin}^{(j)}(t)\right]^2\right) \right]}.
\end{equation}
%----------------------------
Because the occupied rotor subspace is two-dimensional in a fixed-$j$ sector, the I-concurrence coincides with the two-qubit concurrence, ${\cal C}_{I,j}(t) = {\cal C}_j(t)$. Correspondingly, the reduced spin purity is
%----------------------------
\begin{equation}\label{eq:purity_fixed}
{\cal P}_{\rm spin}^{(j)}(t) = \operatorname{Tr} \left( \left[\rho_{\rm spin}^{(j)}(t)\right]^2 \right) = 1-\frac{{\cal C}_j^2(t)}{2}.
\end{equation}
%----------------------------
The corresponding entanglement entropy is
%----------------------------
\begin{widetext}
\begin{equation}\label{eq:entropy_fixed}
S_{\rm ent}^{(j)}(t) = -\left[1-P_j(t)\right] \ln\left[1-P_j(t)\right] - P_j(t)\ln P_j(t).
\end{equation}
\end{widetext}
%----------------------------
Thus, for the pure fixed-$j$ state, the purity, entropy, concurrence, and I-concurrence contain equivalent information about the spin--rotor entanglement.

The concurrence can also be related directly to physical spin--rotor correlations. Using Eq.~\eqref{eq:entangled_state}, one finds
%----------------------------
\begin{equation}\label{eq:angular_variances}
\begin{aligned}
\operatorname{Var}_j(S_z) 		& = P_j(t)\left[1-P_j(t)\right], \\
\operatorname{Var}_j(L_z) 		& = P_j(t)\left[1-P_j(t)\right], \\
\operatorname{Cov}_j(L_z,S_z) 	& = -P_j(t)\left[1-P_j(t)\right],
\end{aligned}
\end{equation}
%----------------------------
where $\operatorname{Cov}_j(L_z,S_z) = \langle L_zS_z\rangle_j - \langle L_z\rangle_j \langle S_z\rangle_j$. Consequently,
%----------------------------
\begin{equation}\label{eq:concurrence_correlations}
\begin{aligned}
{\cal C}_j^2(t) & = 4\,\operatorname{Var}_j(S_z) = 4\,\operatorname{Var}_j(L_z) \\ 
				& =-4\,\operatorname{Cov}_j(L_z,S_z).
\end{aligned}
\end{equation}
%----------------------------
%The concurrence is therefore determined exactly by the fluctuations of either angular-momentum component and by their connected correlation. This relation follows directly from the constraint $L_z+S_z=j$ within a fixed total angular-momentum sector.
\blue{These relations follow directly from the fixed-$j$ constraint $L_z+S_z=j$ and should therefore be regarded as a sector-specific reformulation of the two-level entanglement, rather than as independent general relations between concurrence and angular-momentum correlations.} 

There is also a direct relation to the exchange coherence appearing in the Hamiltonian. In fact,
%----------------------------
\begin{equation}\label{eq:exchange_coherence}
\left\langle S_{+}e^{-i\phi} \right\rangle_j = a_j^*(t)b_j(t), \quad {\cal C}_j(t) = 2\left| \left\langle S_{+}e^{-i\phi} \right\rangle_j \right|.
\end{equation}
%----------------------------
Thus, the spin--rotor concurrence is the modulus of the coherence between the unrecoiled and one-quantum-recoiled branches of the state. Equations~\eqref{eq:concurrence_correlations} and \eqref{eq:exchange_coherence} are exact consequences of the pure fixed-$j$ structure; they should not be interpreted as general identities between entanglement and connected correlations for arbitrary mixed states.

Equation~\eqref{eq:concurrence_fixed} also clarifies the distinction between angular-momentum transfer and entanglement. The concurrence
vanishes for $P_j(t)=0$, when no transfer has occurred, and for $P_j(t)=1$, when the transfer is complete. It reaches its maximum value ${\cal C}_j=1$ at $P_j(t)=1/2$. %Maximum entanglement therefore occurs during the coherent intermediate stage in which the rotor occupies the unrecoiled and recoiled angular-momentum states with equal probabilities, rather than after complete transfer.
\blue{Within the present fixed-$j$ two-level structure, maximal entanglement occurs at the equal-weight point $P_j=1/2$, while the concurrence vanishes at both zero and complete transfer.}

For the arbitrary rotor wave packet considered at the compensation point, the two mechanical branches in Eq.~\eqref{eq:general_compact} need not be orthogonal. Their overlap may be characterized by
%----------------------------
\begin{equation}\label{eq:wavepacket_overlap}
\kappa(t) = \bra{\chi_t}e^{i\phi}\ket{\chi_t}. 
\end{equation}
%----------------------------
Introducing $c_t=\cos(\Delta t/2)$ and $s_t=\sin(\Delta t/2)$, the reduced spin density-matrix obtained from Eq.~\eqref{eq:general_compact} is
%----------------------------
\begin{equation}\label{eq:rho_spin_wavepacket}
\rho_{\rm spin}(t) = 
\begin{pmatrix} 
c_t^2					&	i c_ts_t\,\kappa^*(t) \\
-i c_ts_t\,\kappa(t) 	&	s_t^2
\end{pmatrix}.
\end{equation}
%----------------------------
The rotor now occupies, in general, more than two angular-momentum states, so the appropriate pure-state measure is the I-concurrence.
It is
%----------------------------
\begin{equation}\label{eq:iconcurrence_wavepacket}
{\cal C}_{I}(t) = \left|\sin(\Delta t)\right| \sqrt{1-\left|\kappa(t)\right|^2}, 
\end{equation}
%----------------------------
while the corresponding reduced spin purity is
%----------------------------
\begin{equation}\label{eq:purity_wavepacket}
{\cal P}_{\rm spin}(t) = 1 - \frac{1}{2} \sin^2(\Delta t) \left[ 1-\left|\kappa(t)\right|^2 \right]. 
\end{equation}
%----------------------------
The synchronized one-quantum transfer at the compensation point is therefore independent of the initial rotor wave packet, whereas the
entanglement generated at intermediate times depends on the distinguishability of the unrecoiled state $\ket{\chi_t}$ and the recoiled state $e^{i\phi}\ket{\chi_t}$. For an initial angular-momentum eigenstate, $\kappa(t)=0$, and Eq.~\eqref{eq:iconcurrence_wavepacket} reduces to the fixed-sector result. At the complete-transfer times $t_\pi=(2n+1)\pi/\Delta$, the I-concurrence vanishes for every initial rotor wave packet because the final state again factorizes into spin and rotor components.

%----------------------------------
\begin{figure}[t]
\centering
\includegraphics[width=1.0\linewidth]{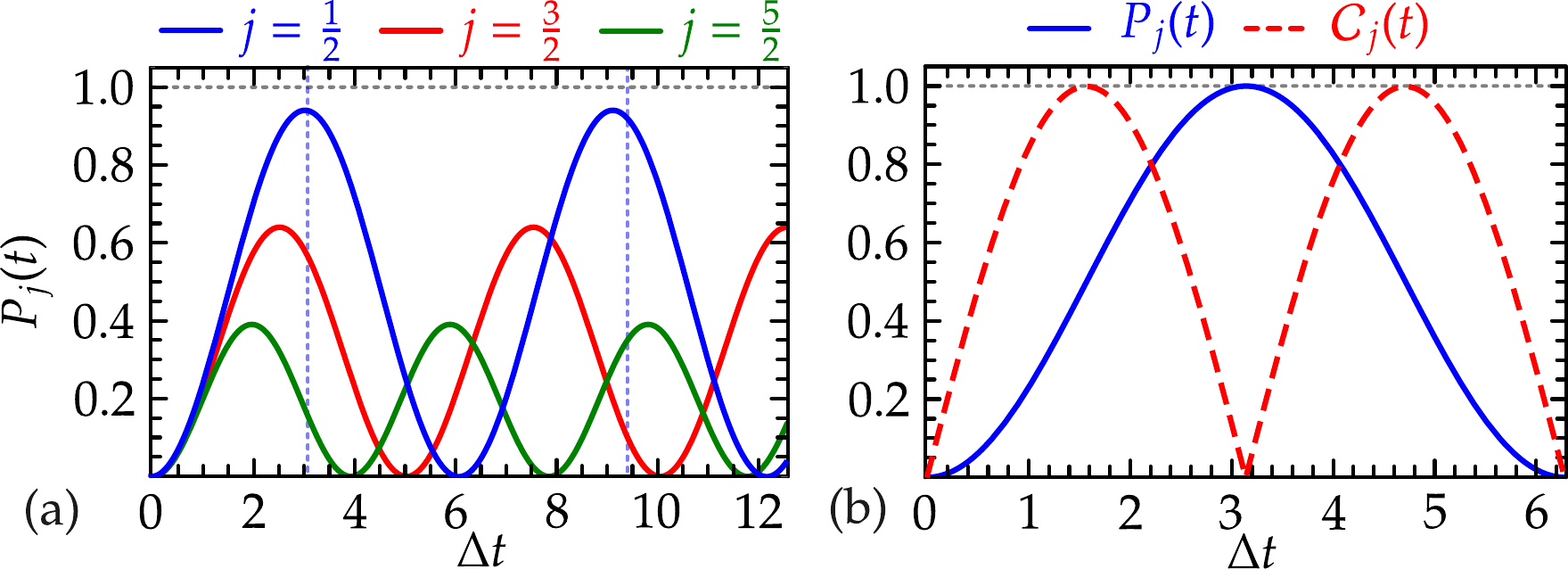}
\caption{(a) Transfer probability $P_j(t)$ for $j=1/2$, $3/2$, and $5/2$ at $\Delta I=1$ and $\dg I=0.5$. The sector-dependent detuning reduces the maximum transfer probability and modifies the oscillation frequency as $|j|$ increases. (b) Transfer probability $P_j(t)$ and concurrence ${\cal C}_j(t)$ for a fixed-$j$ ($j=1/2$) sector at the recoil-compensation point $\dg I=1$. At compensation, $P_j(t)=\sin^2(\Delta t/2)$ and ${\cal C}_j(t)=|\sin(\Delta t)|$. The concurrence reaches its maximal value at half transfer, $P_j(t)=1/2$, and vanishes when the one-quantum angular-momentum transfer is complete at $\Delta t=\pi$.}
\label{fig:Fig2}
\end{figure}
%----------------------------------

%-----------------------------------------------------------------------
\section{Rotor-mediated entanglement of two spins \label{sec.III}}
%-----------------------------------------------------------------------

%----------------------------------
\begin{figure*}[t]
\centering
\includegraphics[width=1.0\linewidth]{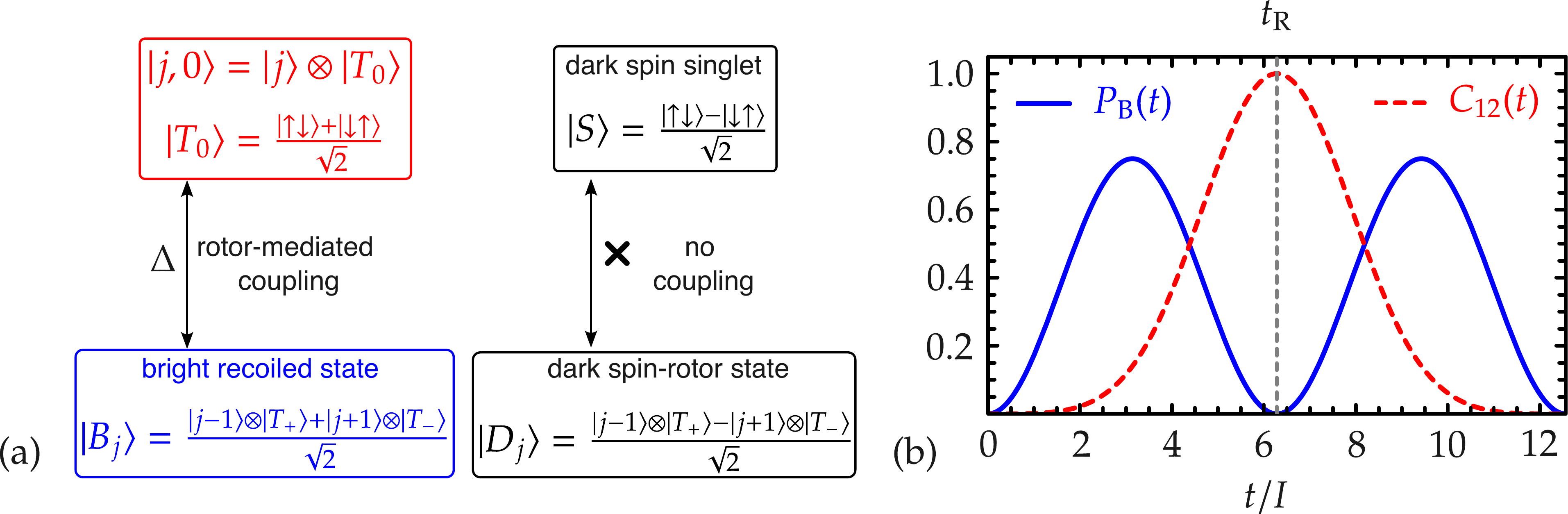}
\caption{(a) Bright- and dark-state structure of the two-spin problem at the recoil-compensation condition $\dg I=1$. The state $\ket{j,0}=\ket{j}\otimes\ket{T_0}$ couples to the bright recoiled state $\ket{B_j}$ with effective coupling $\Delta$, while the singlet $\ket{S}$ and the antisymmetric combination $\ket{D_j}$ remain dark. (b) Bright-state population $P_B(t)$, normalized within the triplet sector, and spin--spin concurrence ${\cal C}_{12}(t)$ at the Bell-state generation point $\dg I=1$ and $\Delta I=\sqrt{3}/4$, starting from the separable spin state $\ket{\uparrow\downarrow}$ and an initial rotor angular-momentum eigenstate. At the first revival time $t_{\rm R}=2\pi I$, the bright-state population vanishes, $P_B(t_{\rm R})=0$, while the concurrence reaches its maximal value ${\cal C}_{12}(t_{\rm R})=1$. The rotor therefore disentangles from the spins and leaves behind a maximally entangled two-spin state.}
\label{fig:Fig3}
\end{figure*}
%----------------------------------

The single-spin problem discussed above shows that the rotor can coherently exchange angular momentum with a localized spin. We now consider a minimal extension to two spin-$1/2$ degrees of freedom coupled to the same rotor. \blue{In contrast to the single-spin problem, this extension introduces a nontrivial collective-spin structure with dark and rotor-active sectors, which is essential for the entanglement-generation mechanism discussed below.} We assume that the two spins couple to the rotor with the same strength, so that the Hamiltonian is
%----------------------------
\begin{equation}\label{eq:two_spin_hamiltonian}
\begin{aligned}
{\cal H}_2 = \frac{L_z^2}{2I} & + \frac{\Delta}{2} \left[ \left(S_{1+}+S_{2+}\right)e^{-i\phi} + \left(S_{1-}+S_{2-}\right)e^{i\phi} \right] \\
							  & + \dg L_z\left(S_{1z}+S_{2z} \right).
\end{aligned}
\end{equation}
%----------------------------
Introducing the total spin operator $\mathbf{S}=\mathbf{S}_1+\mathbf{S}_2$, Eq.~\eqref{eq:two_spin_hamiltonian} has the same form as Eq.~\eqref{eq.1}, with the single spin replaced by the collective spin. Consequently,
%----------------------------
\begin{equation}\label{eq:two_spin_conservation}
[{\cal H}_2,J_z] = 0, \qquad [{\cal H}_2,\mathbf{S}^2] = 0, \qquad J_z = L_z+S_z .
\end{equation}
%----------------------------
The important difference is that the two-spin Hilbert space separates into the singlet sector with $S=0$ and the triplet sector with $S=1$. We use the states %----------------------------
\begin{equation}\label{eq:singlet_triplet_states}
\begin{aligned}
\ket{S} 	& = \frac{1}{\sqrt{2}} \left( \ket{\uparrow\downarrow} -\ket{\downarrow\uparrow} \right), \\
\ket{T_+} 	& = \ket{\uparrow\uparrow}, \\
\ket{T_0} 	& = \frac{1}{\sqrt{2}} \left( \ket{\uparrow\downarrow} +\ket{\downarrow\uparrow} \right), \\
\ket{T_-} 	& = \ket{\downarrow\downarrow}.
\end{aligned}
\end{equation}
%----------------------------
The singlet is decoupled from the spin--rotor exchange because $S_{\pm}\ket{S}=0$ and $S_z\ket{S}=0$. Thus, $\ket{m}\otimes\ket{S}$ is an eigenstate of the full Hamiltonian,
%----------------------------
\begin{equation}\label{eq:singlet_dark}
{\cal H}_2\ket{m}\otimes\ket{S} = \frac{m^2}{2I}\ket{m}\otimes\ket{S}.
\end{equation}
%----------------------------
The singlet therefore forms a dark spin state, while the triplet sector remains coupled to the rotor. \blue{This exact darkness of the singlet relies on the symmetric spin--rotor coupling assumed in Eq.~\eqref{eq:two_spin_hamiltonian}. To see how an asymmetric coupling modifies this structure, let $\Delta_{1,2}=\bar{\Delta}\pm\delta\Delta$ and $g_{1,2}=\bar g\pm\delta g$. The Hamiltonian then acquires the symmetry-breaking contribution
%---------------------------------
\begin{align}
{\cal H}_{\rm asym}
\nonumber
& =
\frac{\delta\Delta}{2}
\left[
(S_{1+}-S_{2+})e^{-i\phi}
+
(S_{1-}-S_{2-})e^{i\phi}
\right] \\
\label{reb.1}
& 
+ \delta g\,L_z(S_{1z}-S_{2z}).
\end{align}
%---------------------------------
These antisymmetric operators couple the singlet and triplet sectors; for example, $(S_{1z}-S_{2z})|S\rangle=|T_0\rangle$ and $(S_{1+}-S_{2+})|S\rangle=-\sqrt{2}|T_+\rangle$. Consequently, for unequal couplings $S^2$ is no longer conserved and the singlet is not an exact dark state, although the conservation of $J_z=L_z+S_z$ is unaffected. Weak coupling asymmetry therefore produces weak singlet--triplet leakage and correspondingly perturbs the ideal revival and entanglement-generation dynamics.}

For the triplet sector, the conserved total angular momentum $j$ is an integer. A fixed-$j$ sector is spanned by the three states
%----------------------------
\begin{equation}\label{eq:triplet_fixed_j_basis}
\ket{j,\pm}	\equiv \ket{j \mp 1} \otimes \ket{T_\pm}, \quad \ket{j,0} 	\equiv \ket{j}\otimes\ket{T_0},
\end{equation}
%----------------------------
In this basis, the triplet Hamiltonian takes the form
%----------------------------
\begin{equation}\label{eq:triplet_matrix}
{\cal H}_{j}^{(T)}
=
\begin{pmatrix}
E_{j,+} 			&	\Delta/\sqrt{2}				&	0					\\
\Delta/\sqrt{2} 	&	E_{j,0} 					&	\Delta/\sqrt{2}		\\
0 					&	\Delta/\sqrt{2} 			&	E_{j,-}
\end{pmatrix},
\end{equation}
%----------------------------
where
%----------------------------
\begin{equation}\label{eq:triplet_diagonal}
E_{j,\pm}	=	\frac{(j \mp 1)^2}{2I} \pm \dg(j \mp 1), \quad E_{j,0} = \frac{j^2}{2I},
\end{equation}
%----------------------------
The factors $\Delta/\sqrt{2}$ follow from the spin-$1$ matrix elements of $S_{\pm}$. Each transition changes the spin angular momentum by one quantum and the rotor angular momentum by the opposite amount, so that $J_z$ remains conserved, as before.

We now return to the recoil-compensation condition $\dg I=1$ identified for the single-spin problem. At this point,
%----------------------------
\begin{equation}\label{eq:triplet_compensation_energies}
E_{j,+} = E_{j,-} = \frac{j^2-1}{2I}, \qquad E_{j,0} = \frac{j^2}{2I}.
\end{equation}
%----------------------------
It is then useful to introduce the symmetric and antisymmetric combinations
%----------------------------
\begin{equation}\label{eq:bright_dark_triplet}
\ket{B_j} = \frac{\ket{j,+}+\ket{j,-}}{\sqrt{2}}, \qquad \ket{D_j} = \frac{\ket{j,+}-\ket{j,-}}{\sqrt{2}}.
\end{equation}
%----------------------------
The state $\ket{D_j}$ is dark, whereas $\ket{j,0}$ couples only to the bright state $\ket{B_j}$. The corresponding $2\times2$ Hamiltonian is
%----------------------------
\begin{equation}\label{eq:two_spin_effective}
{\cal H}_{j}^{\rm eff} = \frac{j^2}{2I}\mathbb{I}_2 + \begin{pmatrix} 
													   0 		&	 \Delta  \\ 
													   \Delta 	& 	-\frac{1}{2I}
													  \end{pmatrix},
\end{equation}
%----------------------------
in the basis $\{\ket{j,0},\ket{B_j}\}$. Apart from the common energy $j^2/(2I)$, this Hamiltonian is independent of $j$. This is the two-spin analogue of the synchronized dynamics found above for a single spin. There is, however, one difference. For the triplet, $S_z^2$ is not a constant. Thus, at $\dg I=1$, the $j$-dependent longitudinal term in Eq.~\eqref{eq.body_frame} is cancelled, but the remaining $-S_z^2/(2I)$ term produces the finite energy difference $1/(2I)$ between $\ket{j,0}$ and $\ket{B_j}$.

The corresponding oscillation frequency is
%----------------------------
\begin{equation}\label{eq:two_spin_rabi_frequency}
\Omega_2 =  \sqrt{4\Delta^2+\frac{1}{4I^2}}.
\end{equation}
%----------------------------
Starting from $\ket{j,0}$, the population of the bright state vanishes again after one complete oscillation at $t_{\rm R}=\frac{2\pi}{\Omega_2}$. At this time the triplet returns to $\ket{j,0}$, while acquiring a phase relative to the dark singlet,
%----------------------------
\begin{equation}\label{eq:two_spin_relative_phase}
\vartheta = \pi\left( 1+\frac{1}{2I\Omega_2} \right).
\end{equation}
%----------------------------

This revival can be used to generate entanglement between the two spins, which was completely absent in the one-spin case discussed before. We consider an initially separable spin state $\ket{\uparrow\downarrow}$ and an arbitrary initial rotor wave packet,
%----------------------------
\begin{equation}\label{eq:two_spin_initial_state}
\ket{\Psi_2(0)} = \ket{\chi}\otimes\ket{\uparrow\downarrow}, \quad \ket{\chi} = \sum_{m\in\mathbb{Z}}c_m\ket{m}.
\end{equation}
%----------------------------
Using $\ket{\uparrow\downarrow} = \frac{1}{\sqrt{2}} \left( \ket{S}+\ket{T_0}\right)$, the dynamics has a simple interpretation. The singlet component remains
dark, while the triplet component undergoes the coherent $\ket{T_0}\leftrightarrow\ket{B_j}$ oscillation. Since Eq.~\eqref{eq:two_spin_effective} is independent of $j$ apart from a common energy shift, all components of the initial rotor wave packet return simultaneously at $t=t_{\rm R}$.

Defining
%----------------------------
\begin{equation}\label{eq:two_spin_rotor_free}
\ket{\chi_t^{(2)}} = \sum_m c_m e^{-i\frac{m^2}{2I}t}\ket{m}, 
\end{equation}
%----------------------------
the state at the first revival can be written as
%----------------------------
\begin{equation}\label{eq:two_spin_revival_state}
\ket{\Psi_2(t_{\rm R})} = \ket{\chi_{t_{\rm R}}^{(2)}} \otimes \frac{1}{\sqrt{2}} \left( \ket{S} + e^{i\vartheta}\ket{T_0} \right).
\end{equation}
%----------------------------
Thus, the rotor is again separated from the spins at the revival time, but the two spins have acquired a relative phase between their singlet and triplet components. To quantify the entanglement between the two spins throughout the evolution, we first trace out the rotor,
%----------------------------
\begin{equation}\label{eq:rho12}
\rho_{12}(t) = {\rm Tr}_{\rm rotor} \left[ \ket{\Psi_2(t)}\bra{\Psi_2(t)} \right].
\end{equation}
%----------------------------
The spin--spin concurrence is then defined by the Wootters formula~\cite{Wootters1998},
%----------------------------
\begin{equation}\label{eq:two_spin_concurrence}
{\cal C}_{12}(t) = \max\left\{ 0,\lambda_1-\lambda_2-\lambda_3-\lambda_4 \right\},
\end{equation}
%----------------------------
where $\lambda_i$ are the square roots of the eigenvalues, in decreasing order, of $\rho_{12}(t)(\sigma_y\otimes\sigma_y)\rho_{12}^{*}(t) (\sigma_y\otimes\sigma_y)$.
At the revival time $t=t_{\rm R}$, the rotor disentangles and the two-spin state becomes pure, so that Eq.~\eqref{eq:two_spin_concurrence} reduces to the usual pure-state expression ${\cal C}_{12}=2|ad-bc|$ for $\ket{\psi}=a\ket{\uparrow\uparrow} + b\ket{\uparrow\downarrow} + c\ket{\downarrow\uparrow} + d\ket{\downarrow\downarrow}$. Applying this to Eq.~\eqref{eq:two_spin_revival_state} gives ${\cal C}_{12}(t_{\rm R})=\left|\sin\vartheta\right|$. A particularly simple point occurs for
%----------------------------
\begin{equation}\label{eq:bell_condition}
\Delta I=\frac{\sqrt{3}}{4}.
\end{equation}
%----------------------------
In this case $\Omega_2=1/I$, so that $t_{\rm R}=2\pi I$ and $\vartheta=3\pi/2$. Consequently,
%----------------------------
\begin{equation}\label{eq:bell_state_generation}
\ket{\Psi_2(t_{\rm R})} = \ket{\chi_{t_{\rm R}}^{(2)}} \otimes \frac{\ket{S}-i\ket{T_0}}{\sqrt{2}}, \quad {\cal C}_{12}(t_{\rm R}) = 1.
\end{equation}
%----------------------------
\blue{The maximally entangled state occurs at a local maximum of the concurrence and is therefore not linearly sensitive to small deviations of the parameter $\Delta I$ or to small timing errors. To see this explicitly, let $x=\Delta I$, so that at the revival time
%------------------------------------
\begin{equation}\label{eq.1d.1}
{\cal C}_{12}(t_{\rm R}) = \left| \sin\left[ \pi\left( 1+\frac{1}{\sqrt{1+16x^2}} \right) \right]\right|.
\end{equation}
%------------------------------------
Expanding around the Bell-state point $x_0=\sqrt{3}/4$ gives
%------------------------------------
\begin{equation}\label{eq.1d.2}
{\cal C}_{12}(t_{\rm R}) = 1-\frac{3\pi^2}{8}(x-x_0)^2 +\mathcal{O}\!\left[(x-x_0)^3\right].   
\end{equation}
%------------------------------------
Thus, the loss of concurrence due to a small deviation of $\Delta I$ from its optimal value is quadratic.

The timing sensitivity can be analyzed similarly. For the initial rotor angular-momentum eigenstate used in Fig.~\ref{fig:Fig3}(b), at $x=x_0$ and $t_{\rm R}=2\pi I$, a small timing error $\delta t$ gives
%------------------------------------
\begin{equation}\label{eq.1d.3}
{\cal C}_{12}(t_{\rm R}+\delta t) = 1-\frac{3}{16} \left(\frac{\delta t}{I}\right)^2 + 
\mathcal{O}\!\left[ \left(\frac{\delta t}{I}\right)^4 \right].
\end{equation}
%------------------------------------
Hence the Bell-state point is a smooth local maximum with respect to both parameter and timing variations, rather than a singular fine-tuned point.}

Up to an overall phase, the two-spin part of this state is
%----------------------------
\begin{equation}\label{eq:bell_state_computational}
\ket{\Psi_{\rm Bell}}
=
\frac{1}{\sqrt{2}}
\left(
\ket{\uparrow\downarrow}
-i\ket{\downarrow\uparrow}
\right).
\end{equation}
%----------------------------
The relative phase $-i$ does not change the degree of entanglement and can be removed by a local rotation about the $z$ axis of either spin; the state is therefore locally equivalent to a conventional Bell state. The dynamics leading to this state is summarized in Fig.~\ref{fig:Fig3}. As shown schematically in Fig.~\ref{fig:Fig3}(a), the singlet component is dark, while the triplet component couples through $\ket{j,0}$ to the bright recoiled state $\ket{B_j}$. The corresponding dynamics is shown in Fig.~\ref{fig:Fig3}(b) for the Bell state-generation point $\dg I=1$ and $\Delta I=\sqrt{3}/4$. The time-dependent concurrence in Fig.~\ref{fig:Fig3}(b) is obtained from the reduced two-spin density matrix after tracing over the rotor. During the evolution, the bright-state population becomes finite, showing the temporary participation of the rotor in the two-spin dynamics. At the first revival time $t_{\rm R}=2\pi I$, however, $P_B(t_{\rm R})=0$, while the spin--spin concurrence reaches ${\cal C}_{12}(t_{\rm R})=1$. Thus, the rotor becomes disentangled from the two spins exactly when maximal spin--spin entanglement is generated. \blue{Experimentally, the condition $P_B(t_{\rm R})=0$ provides a mechanical signature of the revival, while the generated spin coherence can be characterized through phase-sensitive two-spin correlation measurements.} The mechanical degree of freedom therefore acts as a coherent mediator: it participates in the intermediate dynamics, but after one complete cycle it leaves behind a maximally entangled spin pair. \blue{Here, ``mediated'' refers to this dynamical participation of the rotor in the intermediate evolution and the associated singlet--triplet phase accumulation. We do not imply that integrating out the rotor produces
a direct or static effective spin--spin interaction.}

%--------------------------------------------------------------------
\section{Discussion and conclusion \label{sec.IV}}
%--------------------------------------------------------------------

We have introduced an exactly solvable quantum spin--rotor model in which the Barnett and Einstein--de Haas responses follow from the same conservation of total angular momentum. For a single spin-$1/2$, the model gives a particularly simple recoil-compensation point, $\dg I=1$, where all fixed-$J_z$ sectors become synchronized. At this point, an arbitrary initial rotor wave packet undergoes complete spin reversal together with a uniform one-quantum transfer of mechanical angular momentum. The same dynamics also generates spin--rotor entanglement during the intermediate stage of the transfer. \blue{Throughout this work, the terms Barnett and Einstein--de Haas responses are used for their coherent quantum counterparts; recovering the full classical macroscopic dynamics, including environmental relaxation and dissipation,
lies beyond the scope of the present model.}

The two-spin problem reveals a qualitatively different consequence of the same spin--rotor coupling. The Hilbert space separates into a dark singlet and a rotor-active triplet sector, allowing the mechanical degree of freedom to mediate entanglement between two initially separable spins. At $\dg I=1$ and $\Delta I=\sqrt{3}/4$, the rotor completes a coherent cycle and disentangles from the spins, while the two spins are left in a maximally entangled state locally equivalent to a Bell state. Thus, the rotor changes its role from an entangled partner in the single-spin problem to a coherent mediator of spin--spin entanglement in the two-spin problem. This also explains the different relation between population transfer and entanglement in Fig.~\ref{fig:Fig2}(b) and Fig.~\ref{fig:Fig3}(b): while the single-spin concurrence ${\cal C}_j(t)$ is fixed directly by $P_j(t)$, the two-spin concurrence ${\cal C}_{12}(t)$ additionally depends on the relative phase accumulated
between the dark singlet and the rotor-active triplet component. Experimentally, this characteristic dynamics could be identified by combining a measurement of the rotor angular-momentum population with two-spin correlation measurements, for which the disappearance of the bright recoiled population at the revival coincides with maximal spin--spin entanglement.

\blue{An important limitation of the revival protocol is its dependence on the rotor moment of inertia. Restoring $\hbar$, at the Bell-state point the revival time is
%--------------------------------
\begin{equation}\label{eq.1h.1}
t_{\rm R}=\frac{2\pi I_{\rm phys}}{\hbar}.
\end{equation}
%--------------------------------
The corresponding physical timescale therefore depends strongly on the rotor size. As a representative molecular example, Krause \textit{et al.} reported quantized rotational states in the molecular rotor $\mathrm{C_2Sc_2@C_{84}}$ with rotational constant $B=1.73~{\rm cm}^{-1}$~\cite{PhysRevLett.93.137403}, corresponding to $I_{\rm phys}\simeq1.6\times10^{-46}~{\rm kg\,m^2}$. For this inertial scale, Eq.~\eqref{eq.1h.1} gives $t_{\rm R}\simeq9.6~{\rm ps}$. This is many orders of magnitude shorter than spin-coherence times demonstrated in molecular spin systems, which can reach the microsecond regime; for example, Shiddiq \textit{et al.} reported $T_2\simeq8.4~\mu{\rm s}$ \cite{Shiddiq2016}. These examples are not intended to constitute a specific microscopic realization of Eq.~\eqref{eq.1}; rather, they show that the scaling $t_{\rm R}\propto I_{\rm phys}$ does not by itself rule out coherent revival dynamics on molecular rotational scales.

For mesoscopic nanorotors the situation is qualitatively different. Recent spin--rotor proposals based on levitated nanodiamonds involve substantially larger moments of inertia~\cite{zjpt-whpf}. For a representative value $I_{\rm phys}\sim10^{-34}~{\rm kg\,m^2}$, the corresponding free-rotor revival time is of order seconds, and would be far longer than typical spin-coherence times. The present free-rotor revival protocol should therefore not be interpreted as directly applicable to such mesoscopic rotors without additional confinement or control. More generally, observation of the predicted revival requires coherence of the combined spin--rotor system over $t_{\rm R}$; environmental decoherence
and dissipation will reduce the revival contrast and the resulting entanglement. The exactly solvable Hamiltonian studied here describes the coherent limit against which such effects can be assessed.}

These exactly solvable one- and two-spin limits provide a natural foundation for more general spin--rotor systems. With several spins, additional total-spin sectors and spin--spin interactions can produce richer collective dynamics while retaining the same basic competition between spin and mechanical angular momentum. The present results therefore provide simple analytical reference points for studying collective angular-momentum transfer and, in interacting many-spin extensions, more complex quantum correlations such as spin squeezing and multipartite entanglement. \blue{For $N$ spins coupled symmetrically to the same rotor, the interaction can again be expressed in terms of the collective spin, and the construction is similiar to the Dicke model~\cite{PhysRevA.85.043821}; the large-$N$ problem may therefore provide a natural route for investigating the quantum-to-classical crossover of the gyromagnetic dynamics.}

%--------------------------
%\section{Acknowledgments}
%--------------------------

%We acknowledge support from the University of Greifswald, Germany.

%--------------------------------
\bibliographystyle{elsarticle-num}
\bibliography{ref_final}
%--------------------------------

\end{document}